\documentclass[a4paper]{article}

\usepackage[dvipdfmx]{graphicx, color}
\usepackage{wrapfig}
\usepackage{amsmath, amssymb}
\usepackage{bm}
\usepackage{pifont}
\usepackage{cite}
\usepackage[dviout]{pict2e}
\usepackage{booktabs}

\begin{document}

\title{Measuring BR $(h \to \tau ^+ \tau ^-)$ at the ILC: a full simulation study}
\date{March 27, 2015}
\author{Shin-ichi Kawada$^{1,\dagger}$, Keisuke Fujii$^2$, Taikan Suehara$^3$,\\ Tohru Takahashi$^1$, Tomohiko Tanabe$^4$, Harumichi Yokoyama$^5$ }
\maketitle{}

\noindent 1: Graduate School of Advanced Sciences of Matter (AdSM), Hiroshima University, 1-3-1, Kagamiyama, Higashi-Hiroshima, Hiroshima, 739-8530, Japan \\
2: High Energy Accelerator Research Organization (KEK), 1-1, Oho, Tsukuba, Ibaraki, 305-0801, Japan \\
3: Graduate School of Science, Kyushu University, 6-10-1, Hakozaki, Higashi-ku, Fukuoka, 812-8581, Japan \\
4: International Center for Elementary Particle Physics (ICEPP), The University of Tokyo, 7-3-1, Hongo, Bunkyo-ku, Tokyo, 113-0033, Japan \\
5: Graduate School of Science, The University of Tokyo, 7-3-1, Hongo, Bunkyo-ku, Tokyo, 113-0033, Japan \\ \\
$\dagger$ : \verb|s-kawada@huhep.org|

\begin{abstract}
\footnote{Talk presented at the International Workshop on Future Linear Colliders (LCWS14), Belgrade, Serbia, 6-10 October 2014.}
We evaluate the expected measurement accuracy of the branching ratio of the Standard Model Higgs boson decaying into tau lepton pairs at the ILC with a full simulation of the ILD detector concept.
We assume a Higgs mass of 125 GeV, a branching ratio of $\mathrm{BR}(h \to \tau ^+ \tau ^-) = 6.32 \ {\%}$, a beam polarization of electron (positron) of $-0.8 (+0.3)$, and an integrated luminosity of 250 fb$^{-1}$.
The Higgs-strahlung process $e^+ e^- \to Zh$ with $Z\rightarrow q\overline{q}$ is analyzed.
We estimate the measurement accuracy of the branching ratio $\Delta (\sigma \times \mathrm{BR}) / (\sigma \times \mathrm{BR})$ to be 3.4\% with using a multivariate analysis technique.
\end{abstract}

\section{Introduction}

The Higgs boson has been found at the LHC by the ATLAS experiment~\cite{ATLAS} and the CMS experiment~\cite{CMS}.
After that, understanding the properties of a Higgs boson through the precise measurement is important for particle physics.
A branching ratio (BR) is one of the most important properties of a Higgs boson.
In the Standard Model, the Yukawa coupling of matter fermions with a Higgs boson is completely proportional to the fermion mass.
However, the Yukawa coupling will deviate from the prediction of the Standard Model if there are new physics.
The pattern of deviation depends on the new physics model.
Thus, understanding a Higgs boson is very important from the viewpoint of new physics.

Besides, the size of the deviation is expected to be small if the scale of new physics is high.
Specifically, the allowed deviation can be at the few-percent level even if no additional new particles are to be found at the LHC~\cite{deviate}.
Since the branching ratio measurement is used as an input in the extraction of the Yukawa coupling, a precise determination of the branching ratio is essential to probe new physics.

In this study, we focus on the branching ratio of the Higgs boson decaying into tau lepton pairs.
The mass of the tau lepton is known to a very good precision unlike quarks, which typically suffer from the theoretical uncertainties arising from QCD.
Also, the deviation in the lepton Yukawa coupling could well differ from the quark Yukawa coupling, such as in the lepton-specific Two-Higgs Doublet Model.
Thus, the tau Yukawa coupling is an ideal probe for new physics.

In this study, we estimate the measurement accuracy $\Delta (\sigma \times \mathrm{BR}) / (\sigma \times \mathrm{BR})$ of the $h \to \tau ^+ \tau ^-$ branching ratio at the center-of-mass energy of 250 GeV and 500 GeV at the ILC with the ILD full detector simulation.
In this proceedings, we will discuss the 250 GeV case, based on our talk at the LCWS14~\cite{Mytalk}.

\section{Signal and Background}

\begin{figure}[!h]
\centering
\includegraphics[width = 16.0truecm]{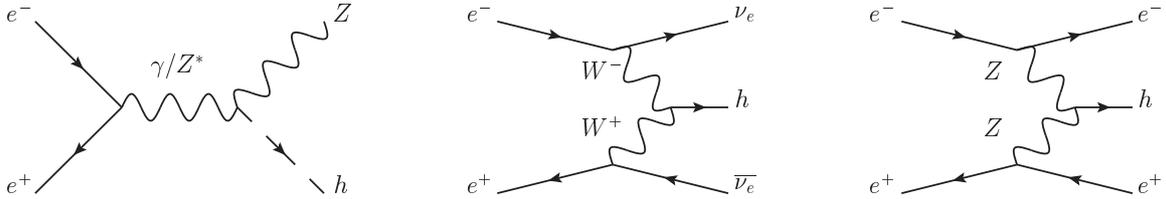}
\caption{The Feynman diagrams of the Higgs boson production processes.
(left): Higgs-strahlung process, (middle): $WW$-fusion process, (right): $ZZ$-fusion process.}
\label{signals}
\end{figure}

The Feynman diagrams of the Higgs boson production processes are shown in Figure~\ref{signals}.
The Higgs-strahlung process ($e^+ e^- \to Zh$) is the most dominant production processes at the center-of-mass energy ($\sqrt{s}$) of 250 GeV.
We analyze this process with the $Z$ boson decaying into quark pair ($Z \to q\overline{q}$), which is expected to be the most sensitive channel because of the high statistics.
The cross section of the Higgs-strahlung process at $\sqrt{s} = 250$ GeV is about 300~fb~\cite{TDR2}.

\clearpage

\begin{figure}[!h]
\centering
\includegraphics[width = 12.0truecm]{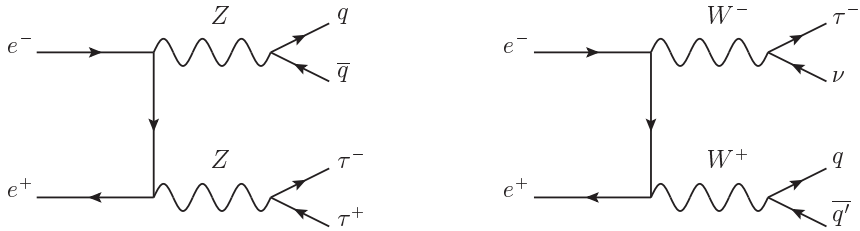}
\caption{The example of Feynman diagrams of the main backgrounds.}
\label{backgrounds}
\end{figure}

The diagrams for the main backgrounds which have the same(similar) final state as the signal are shown in Figure~\ref{backgrounds}.
The $e^+ e^- \to ZZ \to q\overline{q}\tau ^+ \tau ^-$ process shown in left of Figure~\ref{backgrounds} is an irreducible background to the signal.
Besides, the semi-leptonic decay of di-boson processes such as $e^+ e^- \to W^+ W^- \to q\overline{q'}\tau \nu$ shown in Figure~\ref{backgrounds} right will also to be the source of backgrounds due to the similar final states to the signal.

\section{Simulation Conditions}

We assume a Higgs mass of $M_h = 125$ GeV, a branching ratio of $\mathrm{BR}(h \to \tau ^+ \tau ^-) = 6.32 \ \%$~\cite{NNLO}, an integrated luminosity of 250 fb$^{-1}$, and beam polarizations of $P(e^-, e^+) = (-0.8, +0.3)$.

We use the Monte-Carlo samples which are prepared for the studies presented in the ILC Technical Design Report~\cite{TDR2, TDR1, TDR3, TDR4} (so called TDR sample).
Besides, we use the event generator~\cite{CPVHgen} with using GRACE~\cite{GRACE, GRACEweb} and generate Monte-Carlo samples of the processes of $e^+ e^- \to f\overline{f}h$ with $h \to \tau ^+ \tau ^-$ where $f$ denotes a fermion, because the spin correlation of $h \to \tau ^+ \tau ^-$ was not treated properly in TDR samples.
Therefore, we do not use the events which contain $h \to \tau ^+ \tau ^-$ processes in TDR samples to avoid double counting.
The beam energy spectrum includes the effects due to beamstrahlung and the initial state radiation.
The beam-induced backgrounds from $\gamma\gamma$ interactions which give rise to hadrons are included in all signal and background processes.
The background processes from $e^+e^-$ interactions are categorized according to the number of final-state fermions: two fermions (2f), three fermions (1f\verb|_|3f), and four fermions (4f).
We also include $\gamma \gamma \to $~2f and $\gamma \gamma \to $~4f processes for this study.
The detector response is simulated using full simulation based on Geant4~\cite{Geant4}.
We perform the detector simulation with \verb|Mokka|~\cite{Mokka}, a Geant4-based full simulation, with the ILD detector model of \verb|ILD_o1_v05|.
\verb|TAUOLA|~\cite{TAUOLA} is used for the tau decay simulation.
The ILD detector model consists of a vertex detector, a time projection chamber, an electromagnetic calorimeter, a hadronic calorimeter, a return yoke, muon systems, and forward components.

\section{Event Reconstruction}

Even in the signal processes, some hadrons generated via the beam-induced $\gamma \gamma$ interactions are included.
But at the $\sqrt{s} = 250$ GeV, the numbers of hadrons are estimated to be small~\cite{overlay}.
Therefore, we do not apply any special treatment to eliminate these hadrons.

The event reconstruction is consisted of three steps; (1) tau reconstruction, (2) collinear approximation, and (3) $Z$ boson reconstruction.

First, we apply tau finder.
Our tau finder searches for charged track which have highest energy among the remaining particles, and combines the neighboring particles which satisfy $\cos \theta _{\mathrm{cone}} > 0.99$, with the combined mass less than 2 GeV, where $\theta _{\mathrm{cone}}$ is the cone angle with respect to the highest energy charged track.
The combined object is regarded as a tau candidate.
Then, we apply the selection cuts to tau candidates as following; discard the candidates which categorized into 3-prong with neutral particles, $E_{\mathrm{tau \ candidate}} > 3$ GeV where $E_{\mathrm{cand}}$ is the energy of tau candidate, and satisfy $E_{\mathrm{cone}} < ( 0.1\times E_{\mathrm{cand}} )$ with $\cos \theta ' _{\mathrm{cone}} = 0.95$ where $E_{\mathrm{cone}}$ is the cone energy of a tau candidate with the cone angle of $\theta ' _{\mathrm{cone}}$, and $\theta ' _{\mathrm{cone}}$ is the cone angle with respect to a tau candidate, respectively.
These selections are tuned to minimize the misidentification of fragments of quark jets as tau decays.
After the selection, we apply the charge recovery process to obtain better efficiency.
The charged particles in a tau candidate which have the energy less than 2 GeV are detached one by one, the one with the smallest energy first, until satisfying the following conditions; the charge of a tau candidate is exactly equal to $\pm 1$, and the number of track(s) in a tau candidate is exactly equal to 1 or 3.
The tau candidate after the detaching is rejected if it does not satisfy the above conditions, and remaining candidate is regarded as a tau jet.
We repeat the above processes until there are no charged particles which have the energy greater than 2 GeV.

After finishing the tau reconstruction, we apply the collinear approximation~\cite{colapp} to reconstruct the invariant mass of tau pair system.
In this approximation, we assume that the visible decay products of the tau lepton and the neutrino(s) from the tau decay is collinear, and the contribution of the missing transverse momentum comes only from the neutrino(s) from tau decay.

After the approximation, we apply the Durham jet clustering~\cite{Durham} with two jets for the remaining objects to reconstruct $Z$ boson.

\section{Event Selection}

Before optimizing the cuts, we apply preselection cuts as following;
\begin{itemize}
\item the number of reconstructed $\tau ^+$ and $\tau ^-$ are exactly equal to one,
\item the number of reconstructed jet is exactly equal to two,
\item the number of track in an event is greater or equal to nine,
\item $M_{\mathrm{col}} > 0$ and $E_{\mathrm{col}} > 0$ where $M_{\mathrm{col}} (E_{\mathrm{col}})$ is the invariant mass (energy) of tau pair with collinear approximation in the unit of GeV.
\end{itemize}
In addition, we apply the following cuts to suppress the trivial backgrounds;
\begin{itemize}
\item $105 < E_{\mathrm{vis}} < 215$ and $M_{\mathrm{vis}} > 95$, where $E_{\mathrm{vis}} (M_{\mathrm{vis}})$ is the visible energy (mass) in the unit of GeV,
\item $P_t > 40$, where $P_t$ is the sum of the magnitude of the transverse momentum of each particle in the unit of GeV,
\item the thrust in an event should be less than $0.97$,
\item $60 < E_{q\overline{q}} < 175$ and $35 < M_{q\overline{q}} < 160$, where $E_{q\overline{q}} (M_{q\overline{q}})$ is the energy (invariant mass) of reconstructed two jets in the unit of GeV, \item $\cos \theta _{q\overline{q}} < 0.5$, where $\theta _{q\overline{q}}$ is the angle between two jets,
\item $E_{\tau ^+ \tau ^-} < 140$ and $5 < M_{\tau ^+ \tau ^-} < 125$, where $E_{\tau ^+ \tau ^-} (M_{\tau ^+ \tau ^-})$ is the energy (invariant mass) of reconstructed two taus without collinear approximation in the unit of GeV,
\item $\cos \theta _{\tau ^+ \tau ^-} < -0.1$, where $\theta _{\tau ^+ \tau ^-}$ is the angle between two taus,
\item $30 < E_{\mathrm{col}} < 270$ and $15 < M_{\mathrm{col}} < 240$,
\item $65 < M_{\mathrm{recoil}} < 185$, where $M_{\mathrm{recoil}}$ is the recoil mass against $Z$ boson.
\end{itemize}
We only use the events passed the preselection and cuts above for the optimization of event selection.
We perform the cut-based analysis and multivariate analysis both.
The optimization is performed to maximize the signal statistical significance of $S / \sqrt{S+B}$, where S(B) is the number of signal(background).

\subsection{Cut-based Analysis}

\begin{figure}[!h]
\centering
\includegraphics[width = 13.0truecm]{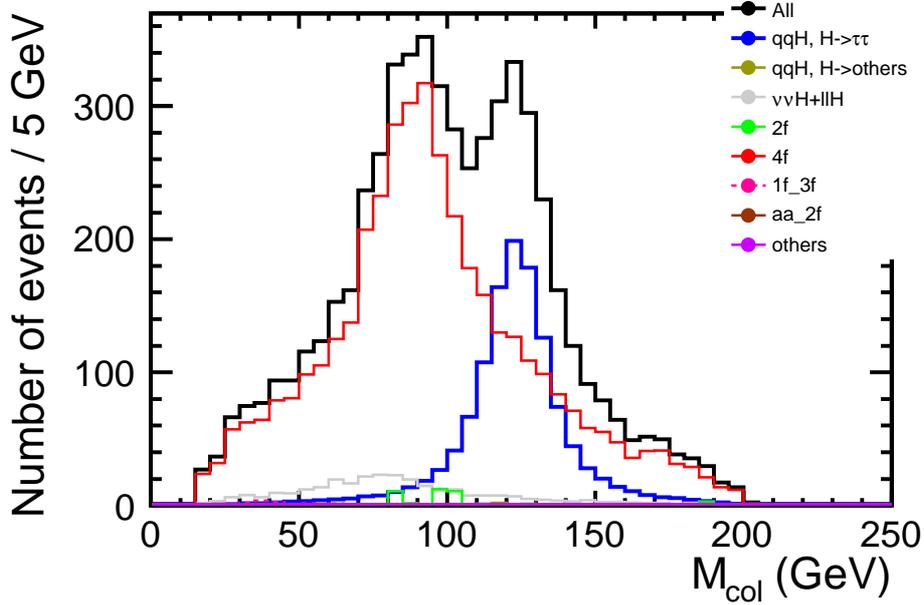}
\caption{The distribution of $M_{\mathrm{col}}$ in the sequential cuts.
Black histogram shows the summing up of all processes, blue shows the signal process, and red shows the all four-fermion background processes, respectively.}
\label{mass_colapp}
\end{figure}

We apply the following cuts sequentially to extract maximum signal significance; 
\begin{itemize}
\item $E_{\mathrm{vis}} < 240$ GeV,
\item $|\cos \theta _{\mathrm{miss}}| < 0.98$, where $\theta _{\mathrm{miss}}$ is the angle between missing momentum in an event and beam axis,
\item $E_{q\overline{q}} < 125$ GeV, $M_{q\overline{q}} > 80$ GeV,
\item $E_{\tau ^+ \tau ^-} < 130$ GeV, $M_{\tau ^+ \tau ^-} < 115$ GeV, $\cos \theta _{\tau ^+ \tau ^-} < 0.54$,
\item $E_{\mathrm{col}} < 210$ GeV, $M_{\mathrm{col}} > 100$ GeV,
\item $\log _{10} |d_0\mathrm{sig}(\tau ^+)| + \log _{10} |d_0\mathrm{sig}(\tau ^-)| > -0.2$, where $d_0\mathrm{sig}(\tau ^+)$ is the impact parameter $d_0$ ($xy$-plane) divided by the error of $d_0$ for $\tau ^+$,
\item $\log _{10} |z_0\mathrm{sig}(\tau ^+)| + \log _{10} |z_0\mathrm{sig}(\tau ^-)| > -0.2$, where $z_0\mathrm{sig}(\tau ^+)$ is the impact parameter $z_0$ ($rz$-plane) divided by the error of $z_0$ for $\tau ^+$,
\item $M_{\mathrm{recoil}} > 115$ GeV.
\end{itemize}
Figure~\ref{mass_colapp} shows the $M_{\mathrm{col}}$ distribution in the sequential cuts.
After all cuts above, the signal events of 1002 and the background events of 535.4 are remained.
The statistical significance is calculated to be $S / \sqrt{S + B} = 25.6 \sigma$.
This result corresponds to the precision of $\Delta (\sigma \times \mathrm{BR}) / (\sigma \times \mathrm{BR}) = 3.9\%$.

\subsection{Multivariate Analysis}

\begin{figure}[!h]
\centering
\includegraphics[width = 13.0truecm]{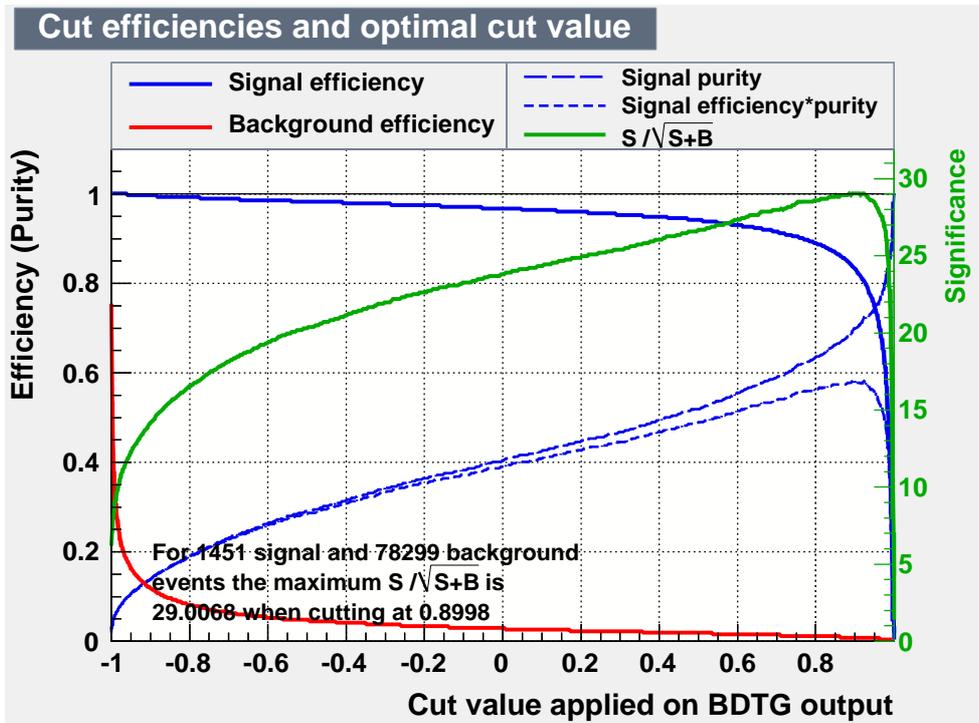}
\caption{The multivariate classifier response for $q \overline{q} h$ mode.
The green plot shows the signal significance.}
\label{BDTG_qqh250}
\end{figure}

We use the TMVA package in ROOT~\cite{ROOT} and use the Gradient Boosted Decision Tree (BDTG) technique for the analysis tool.
We use the following 17 variables as the inputs;
\begin{itemize}
\item three variables from the overall event: $E_{\mathrm{vis}}$, $P_t(\mathrm{all})$, and $\cos \theta _{\mathrm{miss}}$, where $P_t(\mathrm{all})$ is the magnitude of the transverse momentum calculated from the momentum vector of  all visible particles in an event,
\item three variables from reconstructed jets and $Z$ boson: $M_{q\overline{q}}$, $\cos \theta _{q\overline{q}}$, and $\cos \theta _Z$ where $\theta _Z$ is the angle between the momentum of reconstructed $Z$ boson and beam axis,
\item seven variables from reconstructed taus: $M_{\tau ^+ \tau ^-}$, $E_{\tau ^+ \tau ^-}$, $\cos \theta _{\tau ^+ \tau ^-}$, $\cos \theta _{\mathrm{acop}}$, $\cos \theta _H$, $\log _{10}|d_0\mathrm{sig} (\tau ^+)| + \log _{10}|d_0\mathrm{sig} (\tau ^-)|$, $\log _{10}|z_0\mathrm{sig} (\tau ^+)| + \log _{10}|z_0\mathrm{sig} (\tau ^-)|$,  where $\theta _{\mathrm{acop}}$ is the acoplanarity angle between two taus, $\theta _H$ is the angle between the momentum of reconstructed Higgs boson without collinear approximation and beam axis,
\item three variables from collinear approximation: $M_{\mathrm{col}}$, $E_{\mathrm{col}}$, $\cos \theta _{H\mathrm{col}}$ where $\theta _{H\mathrm{col}}$ is the angle between reconstructed Higgs momentum and beam axis with collinear approximation,
\item one from recoil mass: $M_{\mathrm{recoil}}$.
\end{itemize}
Figure~\ref{BDTG_qqh250} shows the response of the multivariate classifier.
The number of events surviving the event selection is 1205 for the signal and 521.4 for the background.
The signal significance is calculated to be $S / \sqrt{S +B} = 29.0\sigma$, which implies a precision of $\Delta (\sigma \times \mathrm{BR}) / (\sigma \times \mathrm{BR}) = 3.4\%$.
The result of the multivariate analysis improved by $\sim 13\%$ than the cut-based analysis.

\clearpage

\section{Comparison with Previous Results}

We previously performed the analysis with fully simulated samples at $\sqrt{s} = 250$ GeV, and the results are written in Ref~\cite{LCNOTE}.
In Ref~\cite{LCNOTE}, the analysis had been performed with the condition of Higgs mass of 120 GeV, and analyzed three signal modes: $q\overline{q}h$, $e^+ e^- h$, and $\mu ^+ \mu ^- h$.
The combined result with the extrapolation to the Higgs mass of 125 GeV was $\Delta (\sigma \times \mathrm{BR}) / (\sigma \times \mathrm{BR}) = 4.2\%$.
The results described in Section 5 are much better than previous study, even the cross section and the branching ratio of $h \to \tau ^+ \tau ^-$~\cite{NNLO} are worse than the case of Higgs mass of 120 GeV and only using $q\overline{q}h$ mode.
These differences mainly come from the mass difference between $Z$ boson and Higgs boson, and the difference of analysis technique.
Now we can apply tighter cuts in the recoil mass against $Z$ boson than previously, because the peak of recoil mass shifted from 120 GeV to 125 GeV.
This difference introduces good separation power between signal and background.
Besides, we only had been performed cut-based analysis in previous.
The cut-based analysis is very simple and easy to understand, but difficult to get better precision.
The multivariate analysis gives us much better results than the cut-based, as described on Section 5.2.

\section{Summary}

We evaluated the expected measurement accuracy of the branching ratio $\Delta (\sigma \times \mathrm{BR}) / (\sigma \times \mathrm{BR})$ of the $h \to \tau ^+ \tau ^-$ mode at $\sqrt{s} = 250$ GeV at the ILC with a full simulation of the ILD detector model, assuming $M_h = 125$ GeV, $\mathrm{BR}(h \to \tau ^+ \tau ^-) = 6.32 \ \%$, $\int L \ dt = 250 \ \mathrm{fb^{-1}}$, and beam polarizations $P(e^-, e^+) = (-0.8, +0.3)$.
We analyzed the $q\overline{q}h$ final state using a cut-based approach and a multivariate approach.
As a result, we expected $\Delta (\sigma \times \mathrm{BR}) / (\sigma \times \mathrm{BR}) = 3.4\%$ with using Gradient Boosted Decision Tree technique.

\end{document}